\documentclass[twoside]{article}
\usepackage{fleqn,espcrc2}
\usepackage{epsfig}

  \title{Solar neutrino oscillations in the quasi-vacuum regime}

  \author{	Daniele Montanino
  \address{	Dipartimento di Scienza dei Materiali 
           	dell'Universit\`a di Lecce,\\
           	Via Arnesano, I-73100 Lecce, Italy}
  \thanks{	email: daniele.montanino@unile.it}
  }%

\begin{document}

\begin{abstract}
Motivated by recent experimental data, we study solar neutrino oscillations in
the range $\delta m^2/E \in [10^{-10},10^{-7}]$ eV$^2$/MeV. In this  range
vacuum oscillations become increasingly affected by (solar and terrestrial)
matter effects for increasing $\delta m^2$, smoothly reaching the MSW regime. A
numerical study of matter effects in such ``quasi-vacuum'' regime is performed.
The results are applied to the analysis of the recent solar neutrino
phenomenology.
\vspace{1pc}
\end{abstract} \maketitle

\section{Introduction}

Flavor oscillations either in vacuum \cite{Po67} or in matter \cite{MSWm}
provide a viable explanation to the solar neutrino problem \cite{NuAs}. The
corresponding survival probability $P_{\nu_e\to\nu_e}$ is determined by the
mass-mixing parameters $\delta m^2$ and $\omega$, as well as by the electron
density profile $N_e(x)$ along the $\nu$ path. 

For solar neutrino oscillations three characteristic lengths can be identified:
the astronomical unit $L=1.496\times 10^8\ {\rm\ km}$, the oscillation
wavelength in vacuum: 
\begin{equation} 
L_{\rm osc} =
\frac{4\pi E}{\delta m^2} = 2.48\times 10^{-3}\frac{\delta m^2}{E}\;{\rm km}\ , 
\label{Losc} 
\end{equation} 
and the the refraction length in matter: 
\begin{equation} 
L_{\rm mat} =
\frac{2\pi}{\sqrt{2}G_F N_e} = \frac{1.62\times 10^4}{N_e}\;{\rm km}\ ,
\label{Lmat} 
\end{equation} 
where $\delta m^2$ is expressed in eV$^2$, the neutrino energy $E$ in MeV and
the electron density $N_e$ in mol$/{\rm cm}^3$. Usually, two asymptotic regimes
can be identified: 1) the ``Mykheyev-Smirnov-Wolfenstein'' (MSW) oscillation
regime \cite{MSWm} ($\delta m^2/E \geq 10^{-7}$ eV$^2$/MeV) characterized by 
$L_{\rm osc}^{\rm MSW} \sim L_{\rm mat} \ll L$, and the ``vacuum'' (VAC)
oscillation regime \cite{Gl87} ($\delta m^2/E\sim O(10^{-11})$ eV$^2$/MeV),
characterized by $L^{\rm VAC}_{\rm osc} \sim L \gg L_{\rm mat}$. In the MSW
regime, the many oscillation cycles in vacuum ($L_{\rm osc}^{\rm MSW}\ll L$)
are responsible for complete decoherence of oscillations, and the survival
probability depends on the detailed density profile in the Sun (and, during
nighttime, in the Earth). Conversely, in the VAC regime the matter effects
suppress oscillations in the Sun and in the Earth, so that (coherent) flavor
oscillations take place only in vacuum, starting approximately from the Sun
surface \cite{Kr85}.

However, there is a transition regime [that might be called of ``quasi-vacuum''
(QV) oscillations], corresponding to the range $10^{-10} \leq \delta m^2/E \leq
10^{-7}$ eV$^2$/MeV, in which oscillations become increasingly affected by
matter effects, and decreasingly decoherent, for increasing values of $\delta
m^2/E$. In this regime, characterized by $L_{\rm mat} \leq  L^{\rm QV}_{\rm
osc} \leq L$, none of the approximations used in the MSW or in the vacuum
regimes can be applied. In the past, QV oscillations have been considered only
in a in few papers (see \cite{Fo00} and reference therein) since typical fits
to solar $\nu$ rates allowed  only marginal solutions in the range where QV
effects are relevant. However, more recent analyses appear to extend the former
ranges of the VAC solutions {\em upwards\/}  and of the MSW solutions {\em
downwards\/}  in $\delta m^2/E$, making them eventually merge in the QV range
(see, e.g., \cite{Go00,Pg00}).

The QVO regime hase been recently studied in detail by Friedland in
\cite{Fr00}. Other numerical \cite{Fo00} and analytical \cite{Pe00}
studies of QVO have also been performed. Here we discuss the results of
\cite{Fo00} in the context of the most recent solar neutrino data.

\section{Calculation of the survival probability}

From general quantum-mechanical arguments we can derive the probability that a
$\nu_e$ produced in the interior of the Sun is detected as a $\nu_e$ on the
Earth \cite{Fo00,Pe88}:
\begin{eqnarray}
P_{\nu_e\to\nu_e}\!\!\!\! &=&\!\!\!\!
P_\odot P_\oplus + (1-P_\odot)(1-P_\oplus)\nonumber\\
                          &+&\!\!\!\! 
2\sqrt{P_\odot(1-P_\odot)P_\oplus(1-P_\oplus)}\, \cos\xi\ ,
\label{Pmat}
\end{eqnarray}
where $P_\odot$ and $P_\oplus$ are the transition probability
$P_{\nu_e\leftrightarrow\nu_1}$ along the two partial paths inside the Sun (up
to its surface) and inside the Earth (up to the detector), and the total phase
$\xi$ can be written as
\begin{equation}
\xi=\frac{\delta m^2}{2E}(L-R_\odot) + \xi_\odot + \xi_\oplus\ ,
\label{Phase}
\end{equation}
where $R_\odot$ is the solar radius and $\xi_\odot$ ($\xi_\oplus$) is the phase
acquired in the Sun (Earth) matter. 

Equation~(\ref{Pmat}) is a general formula that smoothly interpolates from the
vacuum to MSW regime. In particular, in the vacuum limit $P_\odot \simeq
c^2_\omega \simeq P_\oplus$ and $\xi_\odot,\; \xi_\oplus\simeq 0$, we obtain
the standard vacuum formula. Conversely, in the MSW limit the oscillating term
is averaged away and $P_\odot\simeq\sin^2 \omega_m^0 P_c + \cos^2\omega_m^0
(1-P_c)$, where $\omega_m^0$ is the mixing angle at the production point and
$P_c$ is the crossing probability $P_{\nu^m_2\to\nu_1}$ (where $\nu^m_2$ is the
heavier eigenstate in matter), we obtain the standard Parke's formula
\cite{Pa86}.%
\footnote{In all the QV range $\omega_m^0\simeq \pi/2$ so that $P_\odot \simeq
P_c$.}%

he evolution inside the Sun has been performed by direct numerical integration 
of the MSW equation from solar the center to the solar surface.%
\footnote{In \cite{Fo00} has been explicitly verified that the results are
independent from the specific $\nu$ production point,  which can be effectively
taken at the Sun center.}
The electron density was taken from \cite{BaPi}. Figure~2 in \cite{Fo00} shows,
in  the mass-mixing plane, isolines of the difference $c^2_\omega-P_\odot$
(solid  curves), which becomes zero in the vacuum oscillation limit of very
small  $\delta m^2/E$. The variable $\tan^2\omega$ has been chosen in order to
chart the first two octants of the mixing angle range. In \cite{Pe00} a
semianalytical approximation of $P_\odot(\simeq P_c)$ has been find by
modifying the standard resonance prescription \cite{Kr88} with the Maximum
Violation of Adiabaticity (MVA) prescription. This approximation gives
$P_\odot$ with a percent accuracy.

The evolution inside the Earth is done by evolving analytically the MSW 
equations at any given nadir angle $\eta$, using the technique described in 
\cite{Li97}, which is based on a five-step biquadratic approximation of the
density profile from the Preliminary Reference Earth Model (PREM) \cite{PREM} 
and on a first-order perturbative expansion of the neutrino evolution operator.
Figure~4 in \cite{Fo00} shows isolines of the quantity $c^2_\omega-P_\oplus$,
which becomes zero in the vacuum oscillation limit of very small $\delta m^2/E$
for two representative values of the nadir angle.

Regarding the phases $\xi_\odot$ and $\xi_\oplus$ in Eq.~(\ref{Phase}), it has
been shown in \cite{Fo00} that they are negligible for current fits. Only
$\xi_\odot$ could have non-negligible effects in future high-statistics
experiments. A semianalytic formula for the calculation of $\xi_\odot$ has been
recently derived in \cite{Pe00}.

\section{Decoherence of the oscillating term}

For increasing values of $\delta m^2/E$, the oscillating term in
Eq.~(\ref{Pmat}) become increasingly suppressed by energy smearing. The onset
of the corresponding decoherence effect is best studied  for almost
monochromatic neutrinos, namely for the $^7$Be and $pep$ lines, having a width
$\Delta E\sim O({\rm keV})$ \cite{lines}. When $L_{\rm osc}/L$ become greater
of $\Delta E/\langle E\rangle$ (where $\langle E\rangle$ is the average
neutrino energy) the damping effect is important. This happen for $\delta m^2/E
\geq 10^{-8}$ eV$^2$/MeV (and before for continuos spectra).

For narrow line spectra, the energy-averaged oscillating factor can be written
as $\langle \cos\xi\rangle \simeq D(\delta m^2 L/2\langle E\rangle^2) \cos\xi$,
where $D$ is the modulus of the Fourier-transform of the spectrum \cite{Fo00}.
Figure~8 in \cite{Pe00} shows the function $D$ as function of $\delta
m^2/2\langle E\rangle$ for the $^7$Be and $pep$ lines. As anticipated, damping
factor is important for $\delta m^2/E \geq 2\times 10^{-8}$ eV$^2$/MeV. For
higher values of $\delta m^2/E$, the oscillating term can be completely
neglected. 

The survival probability $P_{\nu_e\to\nu_e}$ has to be averaged in time, since
it depends from $t$, both through $P_\oplus$ via $\eta$ and through the phase
$\xi$ via the earth orbit eccentricity. While $P_\oplus(t)$ and $\cos\xi(t)$
can be be easily averaged separately \cite{Fo00,Li97}, the calculation of the
time average of the term $\sqrt{P_\oplus(1-P_\oplus)} \cos\xi$ is in principle
tedious. Fortunately, the Earth effect is non negligible only when the
oscillating term is doubly suppressed ($P_\odot\sim 0$ and $D\sim 0$).
Therefore, in Eq.~(\ref{Pmat}) one can safely take $P_\oplus\simeq c_\omega^2$
as far as the oscillatory term is non negligible.

\begin{figure}[t]
\epsfig{bbllx = 80 , bblly = 200,
	bburx = 500, bbury = 750, width = 7.5truecm,
	figure= 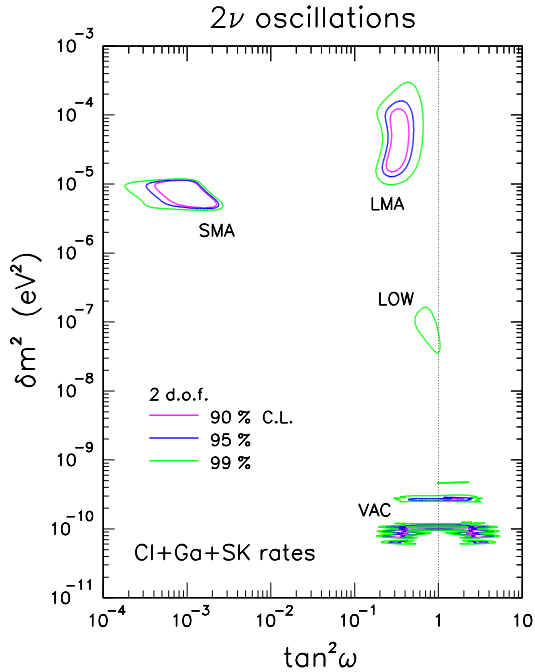}\\
\vspace{-1.5truecm}
\caption{\label{fig:onlyrates} 
         $2\nu$ solutions to the solar neutrino problem at 90, 95 and 99\% 
         C.L.\ of CL+GA+K+SK rates.
        }%
\end{figure}

\begin{figure}[t]
\epsfig{bbllx = 80 , bblly = 200,
	bburx = 500, bbury = 750, width = 7.5truecm,
	figure= 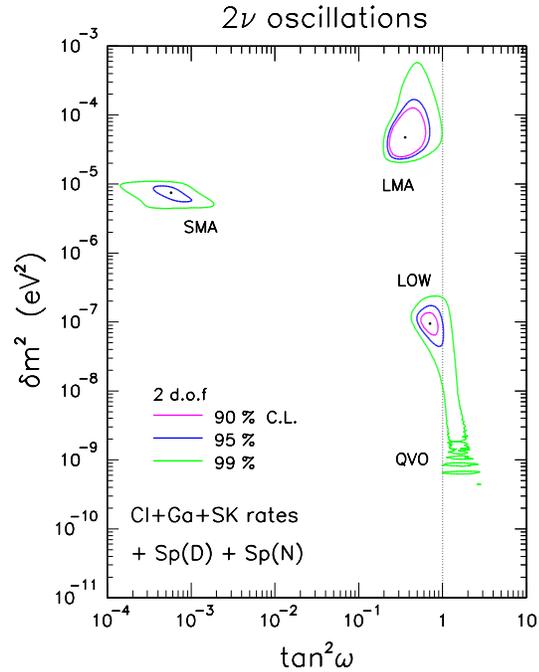}\\
\vspace{-1.5truecm}
\caption{\label{fig:rates+spect} 
         $2\nu$ solutions to the solar neutrino problem  at 90, 95 and 99\% 
         C.L.\ of all data.
         }%
\end{figure}

\section{$2\nu$ solutions to the solar neutrino problem}

In this Section we apply the previous results to the analysis of the more
recent solar neutrino phenomenology. Usually, the analysis is performed in the
first octant in $\omega$ and splitting the $\delta m^2$ range in the vacuum
(${\rm few} \cdot 10^{-11}\leq\delta m^2 \leq {\rm few} \cdot 10^{-10}$ eV$^2$)
and MSW ($\delta m^2 \geq 10^{-8}$ eV$^2$) subranges. 

The lack of observation of a distorsion in the recoil electron spectrum in
SuperKamiokande, together with a rate observed in SuperKamiokande and Gallium
experiments compatible with about $1/2$ suppression of the $\nu_e$ flux,
strongly favor large mixing solutions. The LOW and VAC solutions tend to merge
in the QV part of the plane, and acceptable solutions appear also for
$\omega>\pi/4$ \cite{Go00,Pg00}. Consequently, the correct way to show the
solutions is to use the entire mass-mixing plane without the artificial
splitting in $\delta m^2$ and for $0\leq \omega\leq \pi/2$.

Figure~\ref{fig:onlyrates} shows the combined analysis of the total rate
information coming from the SuperKamiokande (SK) \cite{Su00}, Homestake
\cite{La00}, SAGE \cite{Ga00}, and GALLEX-GNO \cite{Be00} experiments.
Figure~\ref{fig:rates+spect} shows the combination between rate information
and day and night spectral information from SK \cite{Su00}. The technical
details of the analysis are explained in \cite{Fo99} and references therein. 
The value of $\chi^2_{\rm MIN}$ is 35.1 for the LMA, 37.8 for the LOW and 40.7
for the SMA solution, for 36 degrees of freedom.

The LMA and LOW solutions appear to be favored over the SMA solution in the
global fit, although it is rather premature to think that the latter is ruled
out. The vacuum solutions appear strongly disfavored by the nonobservation of
SK spectral distorsions. From Fig.~\ref{fig:rates+spect} we see that a solution
to the solar neutrino problem emerges at 99\% C.L.\ in the QV range. The matter
effects in QV solutions break the symmetry $\omega \to \pi/2-\omega$ valid in
the vacuum range (and that can be well observed in the vacuum solutions in
Fig.~\ref{fig:onlyrates}). From Fig.~\ref{fig:rates+spect} it appears that
large mixing angle solutions are favored. For this reason, several experimental
tests have been proposed to check them \cite{Li00,Pala}.

\section{Conclusions}

We have studied solar neutrino oscillations in the hybrid regime $\delta m^2/E
\in [10^{-10},10^{-7}]$ eV$^2$/MeV (``quasi-vacuum'' regime), in which the
familiar approximations suitable in the vacuum and MSW regime are not valid. We
have discussed a general formula that interpolates smoothly between vacuum and
MSW regimes. The effects of energy smearing on the oscillating term and the
Earth matter effect have been taken in account. The results have been applied
to  the analysis of the most recent solar neutrino data, and suitable solutions
to the solar $\nu$ problem have been found in the QV range.

\section{Acknowledgments}

The author is grateful to G.L.\ Fogli, E.\ Lisi, A.\ Marrone, and A.\ Palazzo
for interesting discussions.



\end{document}